\documentclass[reprint,aps,prl,twocolumn,10pt,superscriptaddress,showpacs]{revtex4-2}
\usepackage{amsmath,amssymb,graphics,epsfig,epstopdf,color,verbatim,ulem,braket,tabularx,multirow,array,diagbox,colortbl,hhline}
\usepackage{blindtext}
\usepackage[colorlinks,linkcolor=blue,citecolor=blue,urlcolor=blue]{hyperref}

\graphicspath{{Figures/}}

\begin{document}

\title{Thermal Entropy, Density Disorder and Antiferromagnetism \\
of Repulsive Fermions in 3D Optical Lattice}

\author{Yu-Feng Song}
\affiliation{Hefei National Laboratory for Physical Sciences at Microscale and Department of Modern Physics, University of Science and Technology of China, Hefei, Anhui 230026, China}
\affiliation{Institute of Modern Physics, Northwest University, Xi'an 710127, China}

\author{Youjin Deng}
\email{yjdeng@ustc.edu.cn}
\affiliation{Hefei National Laboratory for Physical Sciences at Microscale and Department of Modern Physics, University of Science and Technology of China, Hefei, Anhui 230026, China}
\affiliation{Hefei National Laboratory, University of Science and Technology of China, Hefei 230088, China}

\author{Yuan-Yao He}
\email{heyuanyao@nwu.edu.cn}
\affiliation{Institute of Modern Physics, Northwest University, Xi'an 710127, China}
\affiliation{Shaanxi Key Laboratory for Theoretical Physics Frontiers, Xi'an 710127, China}
\affiliation{Hefei National Laboratory, University of Science and Technology of China, Hefei 230088, China}

\begin{abstract}
The celebrated antiferromagnetic (AFM) phase transition was realized in a most recent optical lattice experiment for the 3D fermionic Hubbard model [Shao {\it et al}., Nature {\bf 632}, 267 (2024)]. Despite this important progress, it was observed that the AFM structure factor (and also the critical entropy) reaches the maximum at an interaction strength $U/t\simeq 11.75$, which is significantly larger than the theoretical prediction of $U/t\simeq 8$. Here, we resolve this discrepancy by studying the interplay between the thermal entropy, density disorder, and antiferromagnetism in the half-filled 3D Hubbard model, using numerically exact auxiliary-field quantum Monte Carlo simulations. We have achieved an accurate entropy phase diagram, enabling us to simulate arbitrary entropy path on the temperature-interaction plane and track experimental parameters effectively. We find that above discrepancy can be quantitatively explained by the {\it entropy increase} associated with increasing interaction strength in experiment, and together by the lattice {\it density disorder} present in the experimental setup. We further investigate the entropy dependence of double occupancy and predict universal behaviors that could serve as valuable probes in future optical lattice experiments. 
\end{abstract}

\date{\today}
\maketitle

Quantum simulation combining ultracold atoms with optical lattice potentials~\cite{Anna2007,Bloch2008} has become an essential route to study the intriguing phenomena of strongly correlated systems. A prominent example is the fermionic Hubbard model~\cite{Hubbard1963,Kanamori1963,Gutzwiller1963}, for which optical lattice experiments, equipped with modern cooling, trapping and precision measurement techniques~\cite{Esslinger2010,Christian2017,Florian2020}, enable explorations of broader parameter regimes than those accessible in real materials, such as tunable interaction strengths. Since the early stage, the theoretical interests on the Hubbard model~\cite{Arovas2022,Qin2022} range from quantum magnetism~\cite{LeBlanc2015,Boxiao2017,Xiao2023} and fermionic superfluidity~\cite{Sewer2002,Paiva2010,Fontenele2022} to the more elusive $d$-wave superconductivity~\cite{Qin2020,Haoxu2024}. These remain central pursuits in contemporary optical lattice experiments. Over the past decade, experimental studies of the antiferromagnetic (AFM) spin correlations in the repulsive Hubbard model~\cite{Boll2016,Cheuk2016,Mazurenko2017,Cocchi2017,Muqing2023,Hart2015,Shao2024} have stood in line with or even surpassed the most advanced precision many-body computations~\cite{Qin2022}. 

An inspiring progress in this area is the observation of the AFM phase transition (N\'{e}el transition) in the 3D Hubbard model with ultracold fermionic $^6$Li atoms in a large-scale, nearly uniform optical lattice ($\sim$$800,000$ sites)~\cite{Shao2024}, which has attracted widespread attention~\cite{Niaz2024,Miller2024,Yuanyao2015}. The transition from the paramagnetic (PM) to the N\'{e}el AFM ordered phase was identified via the critical scaling of the measured AFM structure factor, with the critical exponent from the Heisenberg universality class~\cite{Campostrini2002}. However, several key issues remain in the interpretation of this experiment and in its comparison with theoretical predictions. First, the experimental measurements followed irregular paths on the temperature-interaction ($T$-$U$) plane, with entropy as the key variable rather than temperature~\cite{Shao2024}. This complicates direct comparisons with previous many-body numerical calculations of the 3D Hubbard model~\cite{Staudt2000,Paiva2011,Paiva2015,Ibarra2020,Fanjie2024,Song2024a,Song2024b,Khatami2016,Werner2005,Raymond2007,Gorelik2012,Mauro2014,Rampon2024,Kozik2013,Lenihan2022,Garioud2024}, which typically present results as a function of $T$ or $U$. Second, at half-filling, the maximum of the AFM structure factor and the critical entropy appears around $U/t\simeq 11.75$ in the experiment, in contrast to $U/t\simeq 8$ predicted by fixed-temperature calculations using unbiased auxiliary-field quantum Monte Carlo (AFQMC)~\cite{Song2024b}. Understanding this discrepancy and its underlying physics is crucial for advancing quantitative studies of the metal-insulator crossover (MIC) in the PM phase~\cite{Song2024a} and exploring additional phenomena, such as doping effects~\cite{Lenihan2022,Rampon2024}, in the 3D Hubbard model under experimental conditions. Another quantifiable issue in the experiment is the lattice density disorder inherited from the creation of the box trap potential~\cite{Shao2024}. Its impact on various properties of the model, especially the AFM spin correlation, remains to be explored.

In this work, we aim to bridge the gap between the experiment and theory, and resolve the aforementioned discrepancy, by performing numerically exact AFQMC calculations for the half-filled 3D Hubbard model in a manner closely aligned with the experimental setup. Specifically, we use thermal entropy as the central metric, and establish the full map of the entropy per particle ($\boldsymbol{s}$) on the $T$-$U$ plane. To mimic the experimental measurements, we compute the AFM structure factor ($S_{\rm AFM}^{zz}$) along representative entropy paths, observing that the {\it entropy increase} causes the peak location of $S_{\rm AFM}^{zz}$ to shift toward larger $U$. We find that this phenomenon is also contributed by the {\it density disorder}. Additionally, we explore the universal behavior of double occupancy as a function of $\boldsymbol{s}$ with cooling. 

\begin{figure}[t]
\centering
\includegraphics[width=0.995\columnwidth]{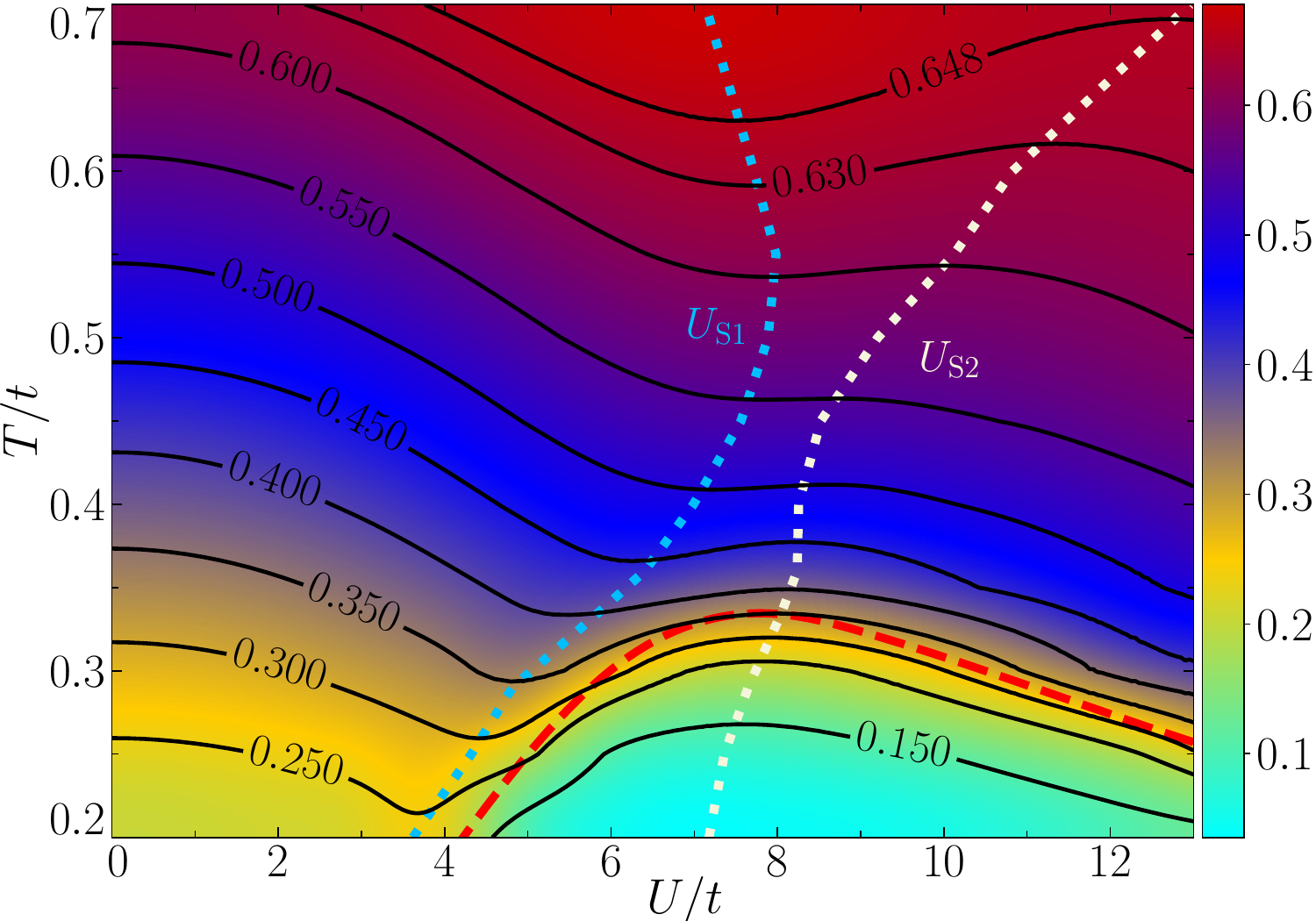}
\caption{\label{fig:Fig01Entropy} Entropy phase diagram of repulsive 3D Hubbard model at half-filling from our AFQMC calculations. Black solid lines are the isentropic curves with the $\boldsymbol{s}$ values (in units of $k_B$) marked on the lines. The residual finite-size effect is negligible. The N\'{e}el temperatures ($T_{N}$, red dashed line), and the positions of local maximum ($U_{\rm S1}$, light blue dotted line) and local minimum ($U_{\rm S2}$, white dotted line) of $\boldsymbol{s}$ at fixed temperatures from Ref.~\cite{Song2024b} are also included. }
\end{figure}

The Hubbard Hamiltonian is $\hat{\mathcal{H}}=\sum_{\mathbf{k}\sigma}\varepsilon_{\mathbf{k}}c_{\mathbf{k}\sigma}^+c_{\mathbf{k}\sigma}^{}+U\sum_{\mathbf{i}}[\hat{n}_{\mathbf{i}\uparrow}\hat{n}_{\mathbf{i}\downarrow}-(\hat{n}_{\mathbf{i}\uparrow}+\hat{n}_{\mathbf{i}\downarrow})/2]+\mu\sum_{\mathbf{i}\sigma}(\hat{n}_{\mathbf{i}\uparrow}+\hat{n}_{\mathbf{i}\downarrow})$, where $\hat{n}_{\mathbf{i}\sigma}=c_{\mathbf{i},\sigma}^+c_{\mathbf{i},\sigma}^{}$ is the density operator with $\mathbf{i}$ labeling a site of the 3D optical lattice and $\sigma=\uparrow,\downarrow$ denoting the two hyperfine states of the ultracold atoms. The kinetic energy dispersion reads $\varepsilon_{\mathbf{k}}=-2t\sum_{\alpha=x,y,z}\cos k_{\alpha}$ with the nearest-neighbor hopping $t$ and the momentum $k_{\alpha}$ defined in units of $2\pi/L$ (the lattice size is $N_s=L^3$). We set $t$ as energy unit, and focus on the model with repulsive interaction $U>0$ and at half-filling ($\mu=0$). The algorithmic and simulation details of the AFQMC calculations can be found in our previous work~\cite{Song2024a,Song2024b}, where we compute the N\'{e}el temperatures and explore the MIC physics in the PM phase of the half-filled 3D Hubbard model from a purely theoretical perspective. In this work, we take a step further to enable our numerical simulations to shake hands with the experiment. 

It is the thermal entropy $\boldsymbol{s}$ instead of temperature which is directly controlled and measured in optical lattice experiments. In numerics, we first compute $\boldsymbol{s}$ versus $U$ for an ensemble of fixed temperatures (with sufficiently small interval)~\cite{Song2024b}, and then we perform an interpolation for the numerical data to achieve the full map of $\boldsymbol{s}$ on the $T$-$U$ plane as presented in Fig.~\ref{fig:Fig01Entropy}. Our results are surely more compact and reliable than the previously reported isentropic properties for half-filled 3D Hubbard model~\cite{Werner2005,Raymond2007}, which applied numerical methods with uncontrolled approximations. As expected, the AFM phase (below the $T_{N}$ line as N\'{e}el transition) has the globally small entropy due to the long-range order. In comparison, $\boldsymbol{s}$ in PM phase exhibits significantly more abundant structures which is closely related to the MIC and effective Heisenberg physics~\cite{Song2024a,Song2024b}. 

The most prominent feature of the entropy map in Fig.~\ref{fig:Fig01Entropy} is the non-monotonic behavior of the isentropic curves $T_i(U)$, along which $\boldsymbol{s}(T_i(U), U)=\boldsymbol{s}_i$ as a constant. With enhancing $U$, The isentropic temperature $T_i(U)$ first decreases, then increases in the middle, and finally decays again. This feature can be understood from the fixed-temperature behavior of $\boldsymbol{s}$ versus $U$ as follows. The total derivative of $\boldsymbol{s}(T_i(U), U)=\boldsymbol{s}_i$ as ${\rm d}\boldsymbol{s}=(\partial\boldsymbol{s}/\partial T){\rm d}T + (\partial\boldsymbol{s}/\partial U){\rm d}U=0$ yields
\begin{equation}\begin{aligned}
\label{eq:DiffEqu0}
\frac{\partial\boldsymbol{s}}{\partial U}\Big|_{T=T_i} = 
-\frac{c(T_i)}{T_i}\frac{{\rm d}T_i(U)}{{\rm d}U},
\end{aligned}\end{equation}
with $c(T)=T\times(\partial\boldsymbol{s}/\partial T)$ denoting the specific heat which is positive for $T>0$. This equality clearly demonstrates that, $T_i(U)$ and $\boldsymbol{s}(T,U)$ have opposite slope signs with respect to $U$ at $T=T_i$, and their extreme points, where ${\rm d}T_i/{\rm d}U=0$ and $(\partial\boldsymbol{s}/\partial U)|_{T=T_i}=0$, should coincide at the same value of $U$. In line with this, it was previously shown~\cite{Song2024b} that, at fixed $T$, $\boldsymbol{s}(T,U)$ initially increases in Fermi liquid regime, reaching a local maximum at $U_{\rm S1}$ (see Fig.~\ref{fig:Fig01Entropy}), then decreases as the system enters the MIC regime, forming a local minimum at $U_{\rm S2}$ (see Fig.~\ref{fig:Fig01Entropy}), and finally increases to $\ln(2)$. This behavior corresponds to the signs $+$, $-$, $+$ for $(\partial\boldsymbol{s}/\partial U)|_{T=T_i}$ in the three regimes separated by $U_{\rm S1}$ and $U_{\rm S2}$. According to Eq.~(\ref{eq:DiffEqu0}), these result in the signs $-$, $+$, $-$ for ${\rm d}T_i/{\rm d}U$ in the respective regimes, thus explaining the shape of the $T_i(U)$ curves. Moreover, we observe that the extreme points of $T_i(U)$ align well with $U_{\rm S1}$ and $U_{\rm S2}$ curves, further confirming the above discussion. 

Remarkably, the interaction-induced adiabatic cooling exists in both the $U<U_{\rm S1}$ and $U>U_{\rm S2}$ regimes, which is evidenced by the decrease of $T$ with increasing $U$ along the isentropic curves plotted in Fig.~\ref{fig:Fig01Entropy}. Nevertheless, this effect was only revealed for the weakly interacting regime (corresponding to $U<U_{\rm S1}$ in our work) of Hubbard models in previous studies~\cite{Werner2005,Raymond2007,Gang2014}. For $U>U_{\rm S2}$, the isentropic curves with $\boldsymbol{s}_i<\ln 2$ should exhibit the asymptotic behavior of $T_i(U)\propto 1/U$ towards $U\to\infty$, as approaching $T_i(U)/J=\alpha$ at which the effective AFM Heisenberg model with the coupling $J=4t^2/U$ has the same entropy. Consequently, the $T_{N}$ line should finally merge to the isentropic curve with $\boldsymbol{s}_i=0.341 k_B$ as the critical entropy of the 3D AFM Heisenberg model~\cite{Wessel2010}, which is again confirmed in Fig.~\ref{fig:Fig01Entropy}. 

\begin{figure}[t]
\centering
\includegraphics[width=0.98\columnwidth]{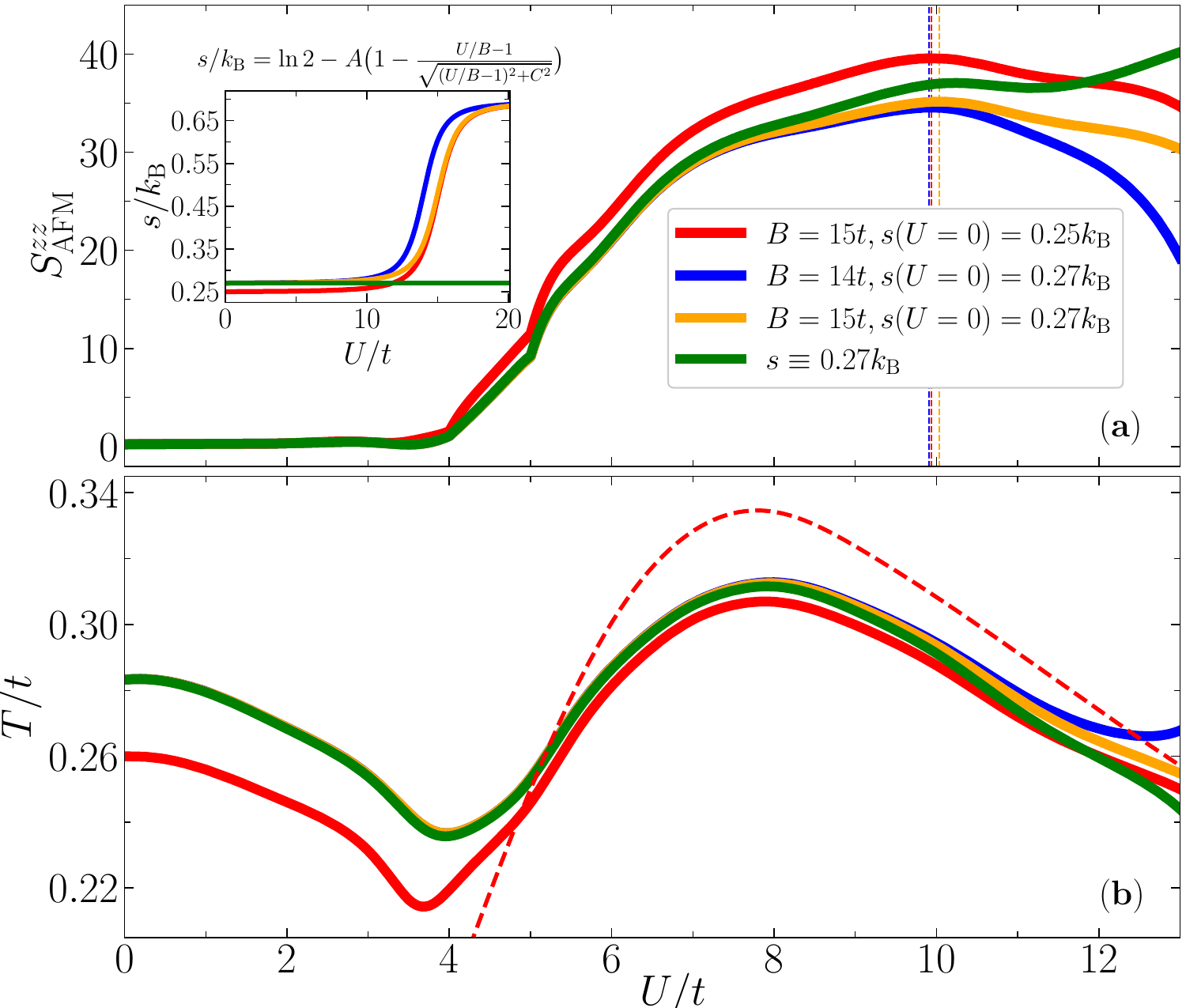}
\caption{\label{fig:Fig02SafmAndT} Numerical results of (a) AFM structure factor $S_{\rm AFM}^{zz}$ and (b) the temperatures $T/t$ as a function of $U/t$, along representative entropy paths. Beside the isentropic path with $\boldsymbol{s}\equiv0.27k_B$ (green solid line), the other three paths (red, blue and yellow solid lines) follow the relation $\boldsymbol{s}/k_B=\ln(2) - A[1-(U/B-1)/\sqrt{(U/B-1)^2+C^2}]$ with three different sets of $A,B,C$ parameters, and are plotted in the inset of (a). The $B$ parameter and the initial entropy $\boldsymbol{s}(U=0)$ for the paths are illustrated in the legend. The vertical dashed lines in panel (a) mark the $S_{\rm AFM}^{zz}$ peak locations, and the N\'{e}el temperatures $T_{N}$ (red dashed line) is included in panel (b). }
\end{figure}

With the above entropy results, we can then take arbitrary entropy path to compute the AFM structure factor defined as $S_{\rm AFM}^{zz}=N_s^{-1}\sum_{\mathbf{ij}}(-1)^{\mathbf{i}+\mathbf{j}}\langle\hat{s}_{\mathbf{i}}^z\hat{s}_{\mathbf{j}}^z\rangle$ with $\hat{s}_{\mathbf{i}}^z=(\hat{n}_{\mathbf{i}\uparrow}-\hat{n}_{\mathbf{i}\downarrow})/2$. (Note the factor of 4 difference in the definition compared to that used in Ref.~\cite{Shao2024}.) The experiment measured $S_{\rm AFM}^{zz}$ versus $U$ along some unknown path crossing the AFM phase on $T$-$U$ plane, and observed a peak at $U_{\rm peak}\simeq 11.75t$. However, previous AFQMC results from fixed-$T$ calculations~\cite{Song2024b} showed that $U_{\rm peak}\simeq 8t$ around the highest N\'{e}el temperature $T_{N}=0.334t$ and it only shifts slightly to a larger value ($<8.6t$) within the temperature range $0.25\le T/t \le0.40$. To understand this disagreement, we first build the $S_{\rm AFM}^{zz}$ map~\cite{Suppl} on the $T$-$U$ plane by applying a similar interpolation scheme to the AFQMC results at various temperatures, which consists of $L=12$ data for $T/t\le0.35$ and converged results (to thermodynamic limit) for $T/t\ge0.40$. 

Then combining the maps of $\boldsymbol{s}$ and $S_{\rm AFM}^{zz}$, we can design specific entropy paths resembling the experiment, and directly read out the corresponding results of $S_{\rm AFM}^{zz}$. This allows us to go beyond the typical scopes of fixed-$T$ or fixed-$U$ calculations. First of all, a special choice is along the isentropic curve meaning adiabatically increasing $U$. As shown in Fig.~\ref{fig:Fig02SafmAndT}(a), for $\boldsymbol{s}(T_i(U), U)\equiv0.27k_B$ as the critical entropy for $U/t=11.75$ (in the sense of initial single-particle entropy at $U=0$) obtained in the experiment~\cite{Shao2024}, $S_{\rm AFM}^{zz}$ is firstly small in Fermi liquid regime, then has a rapid enhancement across the N\'{e}el transition, and finally grows smoothly to the large $U$ limit. Thus, $S_{\rm AFM}^{zz}$ does not even develop a peak versus $U$, as the isentropic curve stays inside the AFM phase [see Fig.~\ref{fig:Fig02SafmAndT}(b)] once entering it and $S_{\rm AFM}^{zz}$ should monotonically increase and converge to the Heisenberg result. This is qualitatively different from the experimental results, and thus confirms the non-adiabatic nature of measurements with increasing $U$. We then choose the specially designed entropy paths depicted by $\boldsymbol{s}/k_B=\ln(2)-A[1-(U/B-1)/\sqrt{(U/B-1)^2+C^2}]$. In this formula, $\boldsymbol{s}/k_B=\ln2$ is guaranteed as the entropy at $U/t=\infty$ (only contributed by the spin degree of freedom), and the initial entropy $\boldsymbol{s}(U=0)$ can be tuned by $(A,B,C)$ parameters with $U=B$ as the fastest increasing location. Considering that $S_{\rm AFM}^{zz}$ almost vanishes at $U/t\simeq 20$ (and thus $\boldsymbol{s}/k_B$ should be close to $\ln2$) and the critical entropy $s_N=0.27k_B$ for $U/t=11.75$~\cite{Shao2024}, we adopt three sets of parameters, one as $B=15t$ and $\boldsymbol{s}(U=0)=0.25k_B$, and the other two as $B=14t$ and $B=15t$ with the same $\boldsymbol{s}(U=0)=0.27k_B$. As plotted in the inset of Fig.~\ref{fig:Fig02SafmAndT}(a), these entropy paths also capture the key experimental features~\cite{Shao2024}: nearly adiabatic evolution at small $U$, strongly non-adiabatic behavior for $U/t>10$, and saturation toward $\ln2$ at $U/t\simeq 20$ as indicated by the vanishing measured $S_{\rm AFM}^{zz}$. Besides, the above formula we adopt for $\boldsymbol{s}/k_B$ is a representative choice that facilitates convenient control of the entropy path. The numerical results of $S_{\rm AFM}^{zz}$ and the temperature curves corresponding to these paths are presented in Fig.~\ref{fig:Fig02SafmAndT}. We can observe that peaks in $S_{\rm AFM}^{zz}$ clearly appear around $U/t\simeq10$ [see Fig.~\ref{fig:Fig02SafmAndT}(a)], which is indeed closer to the experimental observation comparing to fixed-$T$ numerical result of $U_{\rm peak}/t\simeq 8$. Such a peak location shift had not been revealed in previous studies, though it is well known that the experimental path with increasing $U$ typically involves an entropy increase. Moreover, the curve structure and the peak location of $S_{\rm AFM}^{zz}$ depends on $B$ and $\boldsymbol{s}(U=0)$ of the path. The formation of the peak can be attributed to the suppression of $S_{\rm AFM}^{zz}$ due to the re-entrance into PM phase with $U/t>12$ [see Fig.~\ref{fig:Fig02SafmAndT}(b)]. 

We have also tested entropy paths with similar features but described by rather different formulas, and find that the peak properties of $S_{\rm AFM}^{zz}$ and the $\Delta U_{\rm peak}$$\sim$$2t$ shifting as discussed above still exists. Besides, we know that, theoretically, $S_{\rm AFM}^{zz}$ diverges inside AFM phase and converges to a finite value in PM phase. As a result, the $S_{\rm AFM}^{zz}$ peak should become sharper in a large-scale system, as found in experimental results. We emphasize that the {\it entropy increase} encoded in above entropy paths is different from the trivial heating effect, as the actual temperatures for the paths [see Fig.~\ref{fig:Fig02SafmAndT}(b)] can decrease versus $U$ in the regions of $U/t\le4$ and $8\le U/t\le 12$. These numerical results clearly demonstrate that {\it entropy increase} in experimental measurements with increasing $U$ can indeed induce the $S_{\rm AFM}^{zz}$ peak at a larger $U$ value comparing to the fixed-$T$ numerical result. 

\begin{figure}
\centering
\includegraphics[width=0.95\columnwidth]{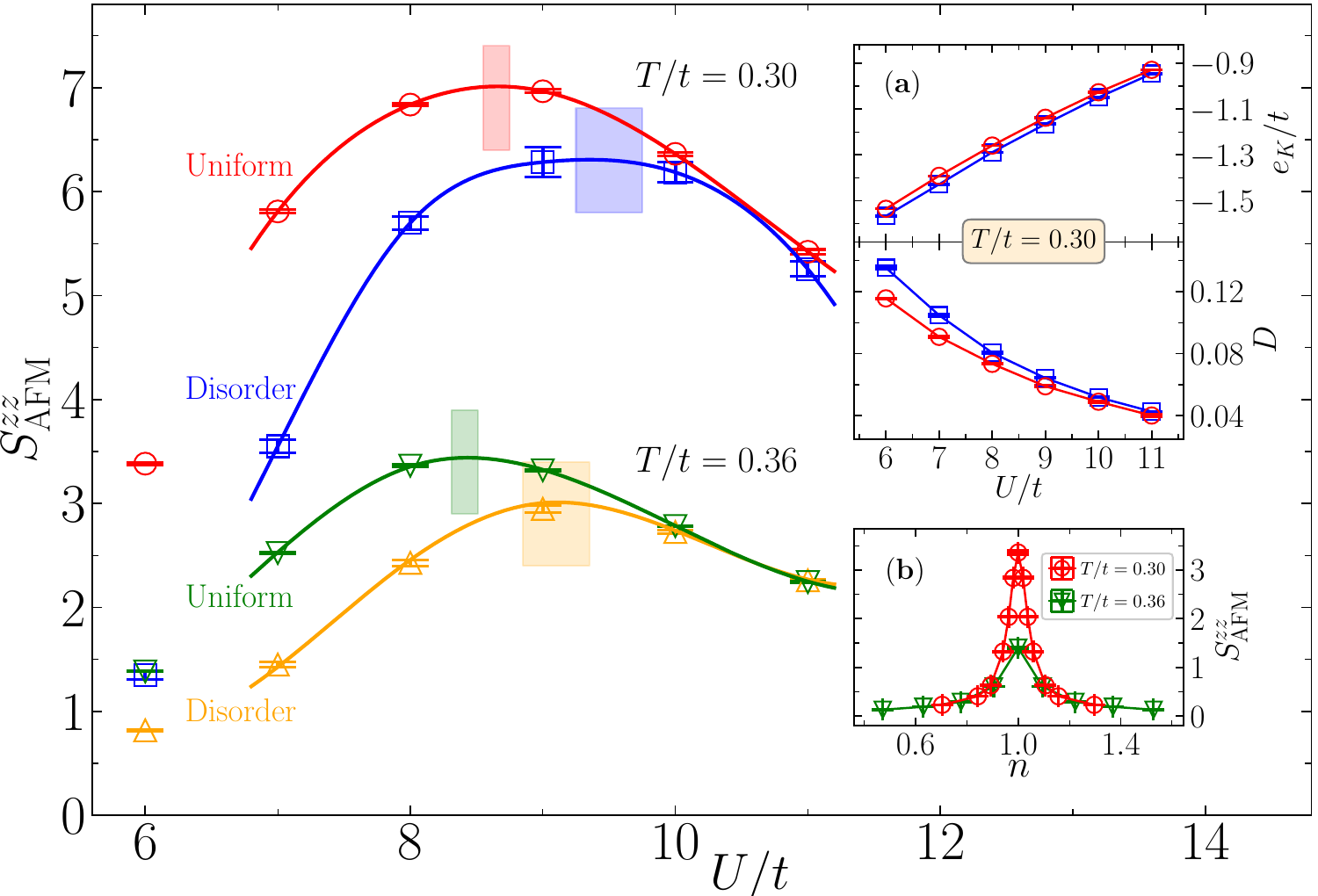}
\caption{\label{fig:Fig03DisOrder} Comparisons of AFM structure factor $S_{\rm AFM}^{zz}$ results versus $U/t$ within periodic boundary conditions (``Uniform'', without disorder) and lattice {\it density disorder} (``Disorder'') for $T/t=0.30$ and $T/t=0.36$. The vertical shading bands denote the peak locations of $S_{\rm AFM}^{zz}$. The inset (a) plots the comparisons of kinetic energy per site $e_K/t$ and double occupancy $D$ between ``Uniform'' and ``Disorder'' calculations for $T/t=0.30$. The inset (b) presents the $S_{\rm AFM}^{zz}$ results from ``Uniform'' versus the fermion filling. These results are from $L=6$ system. }
\end{figure}

We then turn to the lattice {\it density disorder} effect on the AFM structure factor of the model. In AFQMC calculations, we implement a Gaussian disorder by adding a site-dependent chemical potential term $+\sum_{\mathbf{i}}\mu_{\mathbf{i}}(\hat{n}_{\mathbf{i}\uparrow}+\hat{n}_{\mathbf{i}\downarrow})$ to the Hamiltonian, with $\mu_{\mathbf{i}}$ generated by sampling with the normal distribution $P(\mu_{\mathbf{i}})=e^{-\mu_{\mathbf{i}}^2/2}/\sqrt{2\pi}$ with $\bar{\mu}=0$ and $\sigma_{\mu}=t$ as the average and standard deviation. Such calculations guarantee the overall half-filling condition, and also render the normal distribution feature for the lattice density results. Thus, it should be more reasonable and fits better to the experiment~\cite{Shao2024}, than the uniform disorder adopted in a previous study~\cite{Paiva2015}. The value of $\sigma_{\mu}=t$ is chosen such that the resulting local density distribution from our simulations resembles that observed in experiments at comparable temperatures. For every set of parameters, we find that $\sim$$10$ disorder realizations are sufficient to reach the converged disorder average results of AFM spin correlations and $S_{\rm AFM}^{zz}$. The simulation with above disorder is unfortunately limited to $L=6$ system due to fermion sign problem. 

In Fig.~\ref{fig:Fig03DisOrder}, we present the comparisons of $S_{\rm AFM}^{zz}$ versus $U/t$ from AFQMC calculations with (``Disorder'') and without (``Uniform'') {\it density disorder} at $T/t=0.30$ and $0.36$. We can observe that the $S_{\rm AFM}^{zz}$ peak possesses a $\Delta U_{\rm peak}$$\sim$$t$ shifting, which tends to be larger at lower temperature. This shifting can be understood from the fact that, the disorder typically suppresses the correlation length of AFM spin correlation, and thus reduces the $S_{\rm AFM}^{zz}$ result, which is generally more prominent in the small to intermediate $U$ regimes. With strong interaction, the system evolves into a Mott insulator state~\cite{Song2024b} which is almost immune to the disorder effect (given $\mu_i$ is smaller than the gap). As a result, a stronger {\it density disorder} can produce a larger shift of $U_{\rm peak}$. Moreover, the lower kinetic energy and more double occupancy with disorder [inset (a) of Fig.~\ref{fig:Fig03DisOrder}] also reveal that the disordered system is more itinerant than the uniform system. These results together suggest that the {\it density disorder} effectively weakens the interaction of the system. As a result, density disorder can suppress the N\'{e}el transition temperatures in the weak to intermediate interaction regime, thereby shifting the entire AFM ordered phase toward larger values of $U$. This effect may account for the residual deviation between our results in Fig.~\ref{fig:Fig02SafmAndT}(b) and the experimental observation regarding the critical $U$ at which the system enters the AFM ordered phase following the entropy paths. Thus, it is important to consider this issue in the benchmark between experiment and theory. The {\it density disorder} should also be responsible for the relatively small $S_{\rm AFM}^{zz}$ results as measured in Ref.~\cite{Shao2024} especially at low temperatures, comparing to results of perfectly uniform system. For example, the finite-size scaling analysis~\cite{Song2024b} yielded the relation $S_{\rm AFM}^{zz}L^{\eta-2}\simeq 0.2$ (with $\eta=0.0375$) at N\'{e}el transition for $U/t=10$. This predicts $S_{\rm AFM}^{zz}\simeq 1454$ for the uniform system with the same size as that in experiment, comparing to the maximal $S_{\rm AFM}^{zz}\simeq 30$ measured for $U/t=11.75$ inside AFM phase. (More recently, with further improvement of the experimental system, the maximum of $S_{\rm AFM}^{zz}$ has been significantly enhanced up to $\sim$$150$~\cite{CommNote}.) This significant suppression of $S_{\rm AFM}^{zz}$ can be understood from the quantitative equivalence between adding the disorder and locally doping the uniform system. For the latter, $S_{\rm AFM}^{zz}$ rapidly decays away from half filling within $L=6$ [inset (b) of Fig.~\ref{fig:Fig03DisOrder}], which should be even more significant with increasing $L$.

\begin{figure}[t]
\centering
\includegraphics[width=0.95\columnwidth]{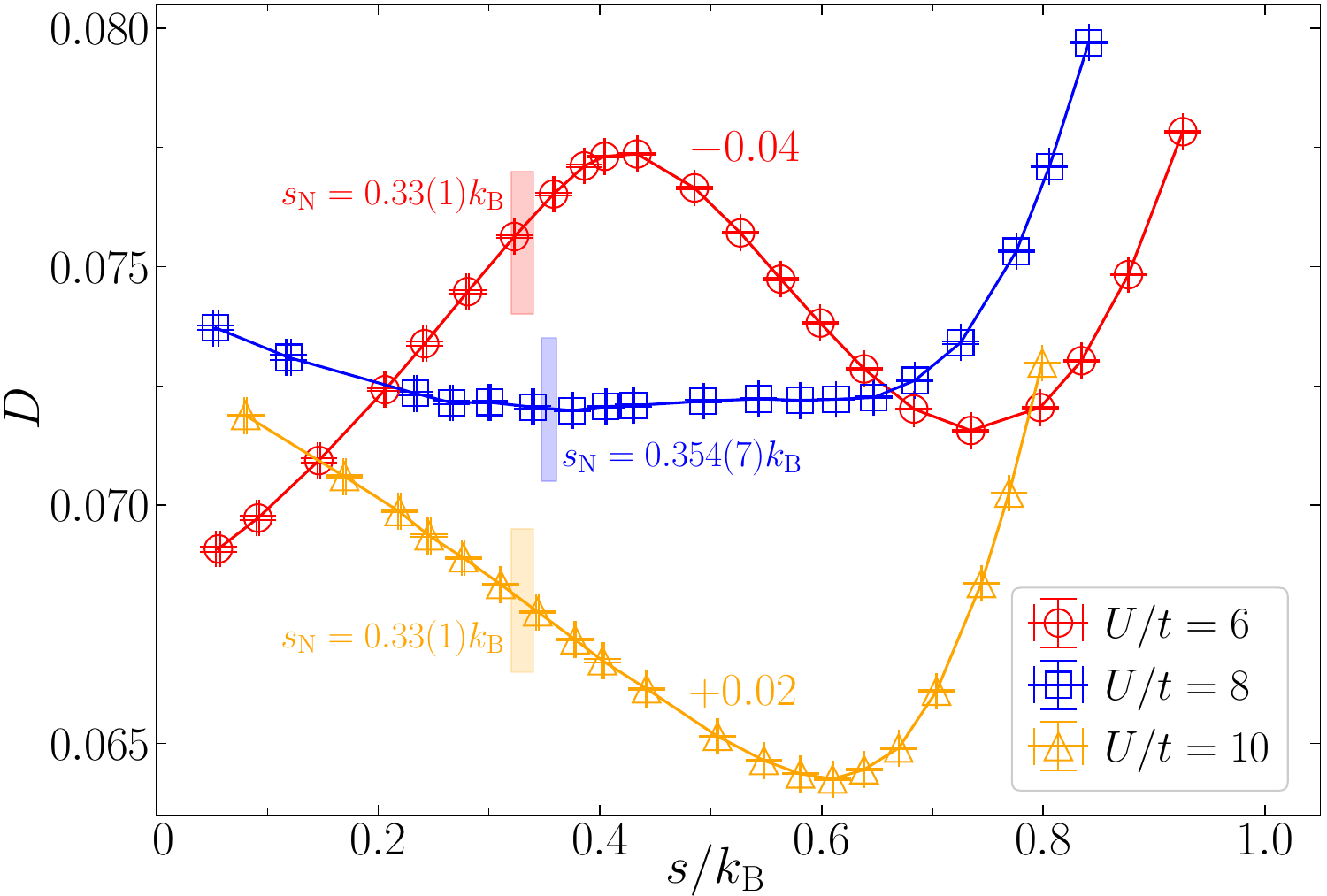}
\caption{\label{fig:Fig04DouOcc} Double occupancy $D$ as a function of the entropy per particle $\boldsymbol{s}$ for $U/t=6,8$ and $10$. For $U/t=6$ and $U/t=10$, the results are shifted by $-0.04$ and $+0.02$ for the plot. The corresponding critical entropy values with uncertainties are marked by the vertical shading bands. The residual finite-size effect is negligible. }
\end{figure}

Computing physical observables as a function of the entropy $\boldsymbol{s}$ instead of temperature is crucial for the benchmark with optical lattice experiment~\cite{Gorelik2012,Shao2024}. Here we report AFQMC results of the double occupancy $D=N_s^{-1}\sum_{\mathbf{i}}\langle\hat{n}_{\mathbf{i}\uparrow}\hat{n}_{\mathbf{i}\downarrow}\rangle$ versus $\boldsymbol{s}$, as presented in Fig.~\ref{fig:Fig04DouOcc}. All the curves $D(\boldsymbol{s})$ should start from the special point ($D=1/4,\boldsymbol{s}/k_B=2\ln2$) as the high-$T$ limit. Upon cooling, we can observe quite different behaviors as well as some universal signatures in $D(\boldsymbol{s})$ for $U/t=6,8,10$, as representatives of weak, intermediate and strong interactions. First, the local maximum/minimum in $D(\boldsymbol{s})$ is actually connected to $U_{\rm S1}$ and $U_{\rm S2}$ in Fig.~\ref{fig:Fig01Entropy}: the equality $(\partial D/\partial\boldsymbol{s})_U=(\partial D/\partial T)_U\times [(\partial\boldsymbol{s}/\partial T)_U]^{-1}$ and the Maxwell’s relation $-(\partial D/\partial T)_{U}=(\partial\boldsymbol{s}/\partial U)_{T}$ explicitly state that $(\partial D/\partial\boldsymbol{s})_U=0$ and $(\partial\boldsymbol{s}/\partial U)_{T}=0$ coincide in $T$-$U$ plane. The $U_{\rm S1}$ curve crosses the vertical $U/t=6$ line for twice (the cross at $T/t$$\sim$$0.8$ is not shown), in correspondence to the local minimum (at $\boldsymbol{s}/k_B\simeq0.72$) and maximum (at $\boldsymbol{s}/k_B\simeq0.40$) of $D(\boldsymbol{s})$ for $U/t=6$ in Fig.~\ref{fig:Fig04DouOcc}. Physically, these local extrema signify the crossovers from bad metal state to Fermi liquid regime and the reverse~\cite{Song2024a}, respectively. In contrast, the situation for strong interactions is quite different. The local minimum at $\boldsymbol{s}_{\rm min}/k_B\simeq0.61$ in $D(\boldsymbol{s})$ for $U/t=10$ corresponds to $U_{\rm S2}$, and this $\boldsymbol{s}_{\rm min}/k_B$ should gradually approach the universal number $\ln 2$ as $U/t\to\infty$~\cite{Gorelik2012}. In this case, the local minimum separates two distinct regions dominated by the charge physics ($\boldsymbol{s}>\boldsymbol{s}_{\rm min}$) and the spin-exchange physics ($\boldsymbol{s}<\boldsymbol{s}_{\rm min}$), and the latter can be effectively described by the AFM Heisenberg model. For $U/t=8$, we observe a plateau within $0.25<\boldsymbol{s}/k_B<0.65$ in $D(\boldsymbol{s})$. The other important feature of the $D(\boldsymbol{s})$ curves is the linear behavior in a low-entropy region containing N\'{e}el transitions, especially for $U/t=6$ and $10$. We have also calculated the critical entropy $\boldsymbol{s}_{\rm N}/k_B=0.33(1)$, $0.354(7)$ and $0.33(1)$ for the three interactions, where the $U/t=10$ result is already quite close to that of the 3D AFM Heisenberg model~\cite{Wessel2010}. These properties in $D(\boldsymbol{s})$ can serve as useful probes in optical lattice experiments to detect the MIC in PM phase, effective Heisenberg physics and the N\'{e}el transition simply by the double occupancy. We have also checked the results of $S_{\rm AFM}^{zz}$ versus the entropy~\cite{Suppl}, and confirmed similar critical scaling behavior $(S_{\rm AFM}^{zz}-S_0)\propto|\boldsymbol{s}/\boldsymbol{s}_{\rm N}-1|^{-\gamma}$ as observed in experiment. 

In summary, we have systematically studied the interplay between the thermal entropy, density disorder and AFM properties in half-filled 3D Hubbard model following the manner of the most recent optical lattice experiment~\cite{Shao2024}. We have established the full entropy map on $T$-$U$ plane for the model, based on which we have resolved the discrepancy between the experiment and theory on the peak antiferromagnetism. More specifically, we have found that both the {\it entropy increase} in experimental measurements during increasing interaction and {\it density disorder} induce the peak location shifting of AFM structure factor towards stronger interactions. We have also presented predictions for the universal behaviors of double occupancy versus entropy to probe various properties of the system, which can serve as useful guidance for future experimental studies. Our work bridges the gap between recent experiment and cutting-edge quantum many-body computations, and establishes a paradigm for benchmarking between them, by taking the entropy as a central role and including the density disorder as a realistic issue. This paradigm is important and applicable to the benchmark studies on many other exotic properties of various versions of the Hubbard model~\cite{Arovas2022,Qin2022} between experiment and theory. 

{\it{Acknowledgements}}. We thank Xing-Can Yao for valuable discussions on the experimental details. This work was supported by the National Natural Science Foundation of China (under Grants No. 12247103, No. 12204377, and No. 12275263), the Innovation Program for Quantum Science and Technology (under Grant No. 2021ZD0301900), the Natural Science Foundation of Fujian province of China (under Grant No. 2023J02032), and the Youth Innovation Team of Shaanxi Universities.

\bibliography{3DHubbEntropyAFMRef}


\clearpage
\onecolumngrid

\begin{center}
\textbf{\large Supplementary material for ``Thermal Entropy, Density Disorder and Antiferromagnetism of Repulsive Fermions in 3D Optical Lattice''}
\end{center}

\section{The heat maps of AFM structure factor and double occupancy}
\label{sec:HeatMap}

\begin{figure}[h]
\centering
\includegraphics[width=0.485\columnwidth]{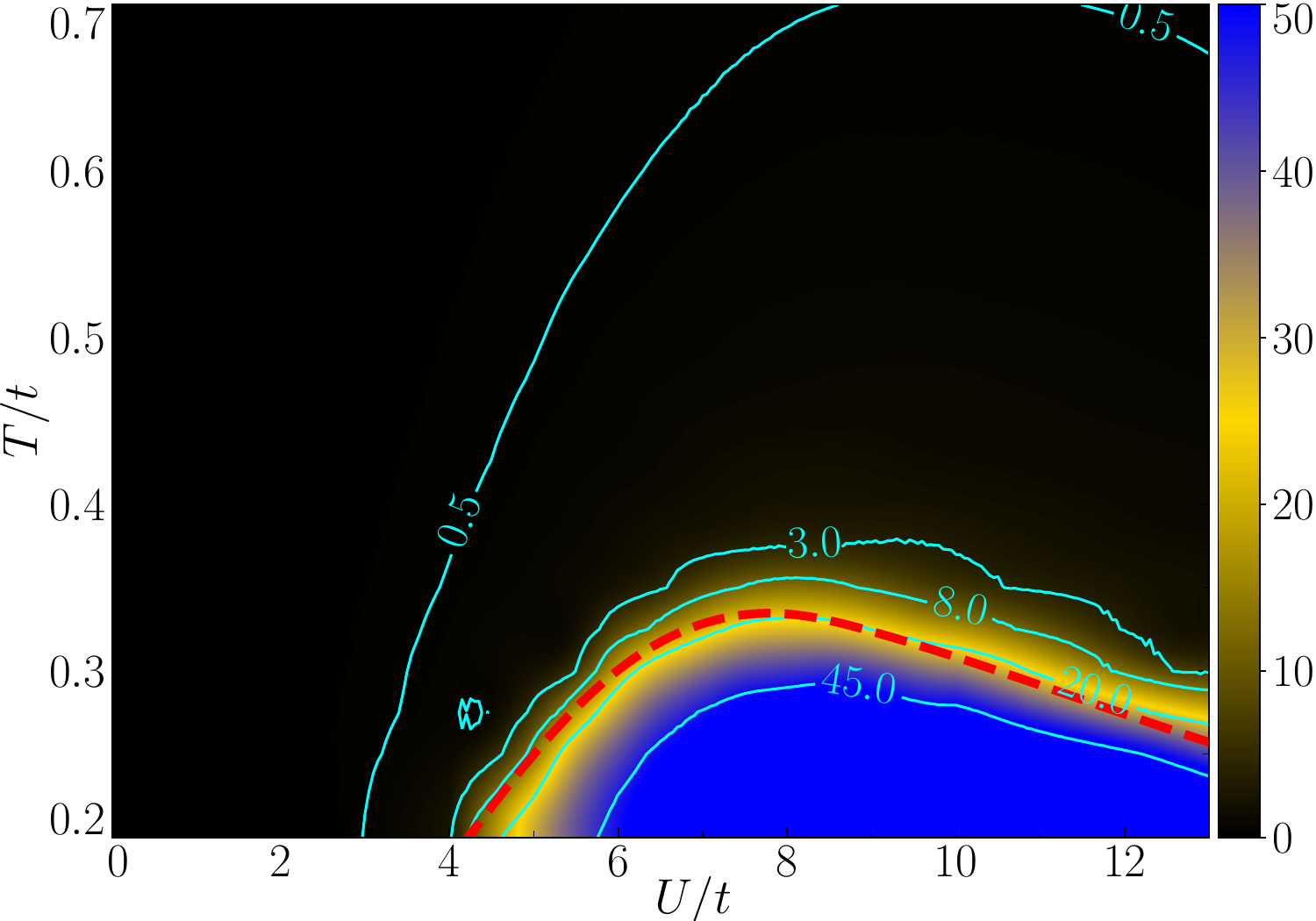}
\hspace{0.2cm}
\includegraphics[width=0.485\columnwidth]{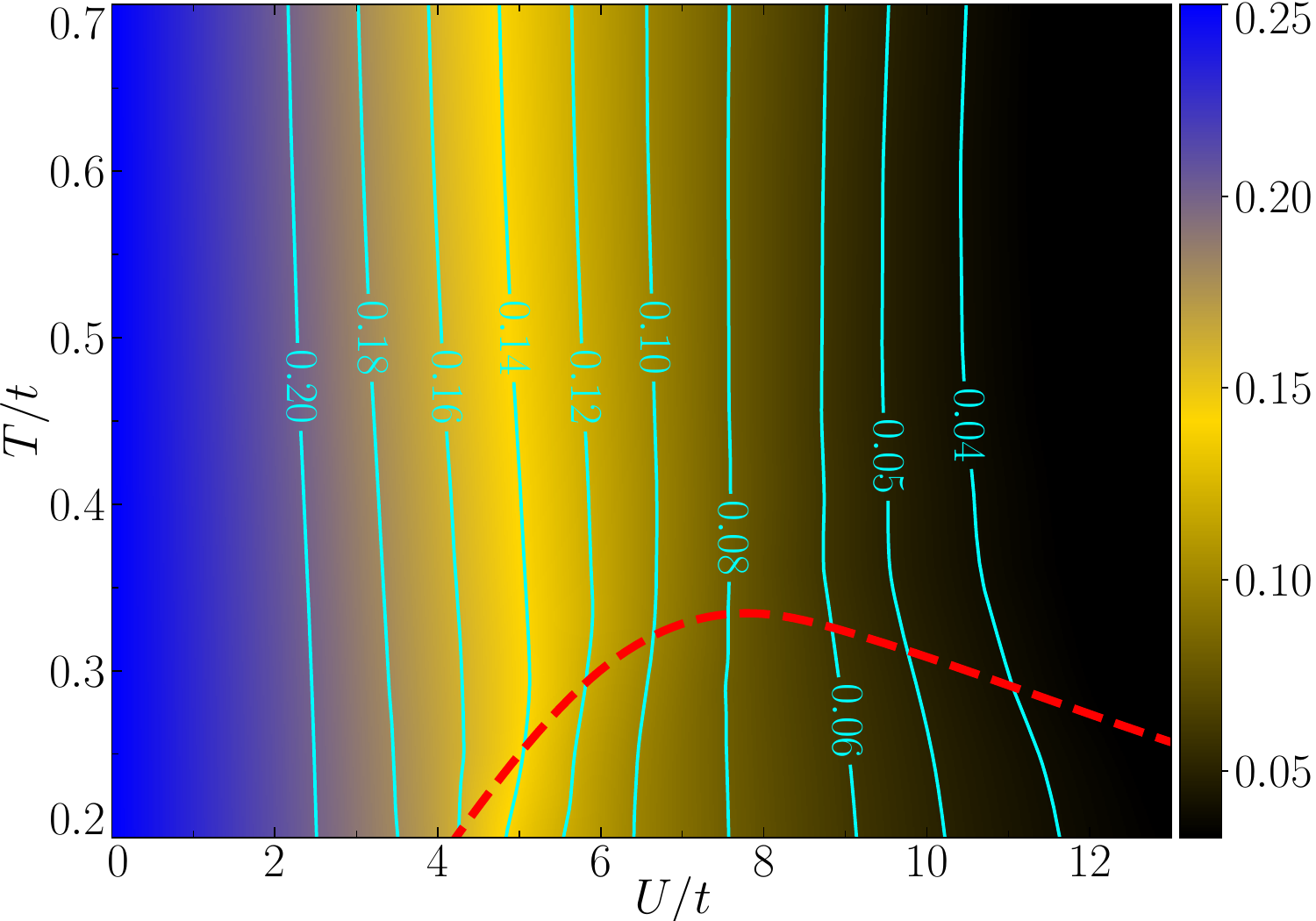}
\caption{\label{fig:FigS1HeatMap} The heat map of AFM structure factor $S_{\rm AFM}^{zz}$ and double occupancy $D$ on $T$-$U$ plane for 3D half-filled Hubbard model from our AFQMC simulations. For $S_{\rm AFM}^{zz}$, the results consist of $L=12$ data for $T\le0.35$ and converged results (to thermodynamic limit) for $T\ge0.40$. For $D$, the residual finite-size effect is negligible. The red dashed line plots the N\'{e}el temperatures $T_{N}$. }
\end{figure}

In Fig.~\ref{fig:FigS1HeatMap}, we present the numerical results of the full maps of antiferromagnetic (AFM) structure factor $S_{\rm AFM}^{zz}$ and double occupancy $D$ on $T$-$U$ plane for 3D half-filled Hubbard model. As expected, the finite-size $S_{\rm AFM}^{zz}$ result is much bigger inside the AFM ordered phase. Note that $S_{\rm AFM}^{zz}$ should diverge in AFM phase and gradually saturate to finite values in paramagnetic (PM) phase as approaching $L=\infty$. The contour lines of double occupancy reveal that this quantity has very weak temperature dependence, especially in PM phase. Moreover, the shape of contour lines of $D$ bends a little bit when crossing the N\'{e}el transitions. 

\begin{figure}[h]
\centering
\includegraphics[width=0.60\columnwidth]{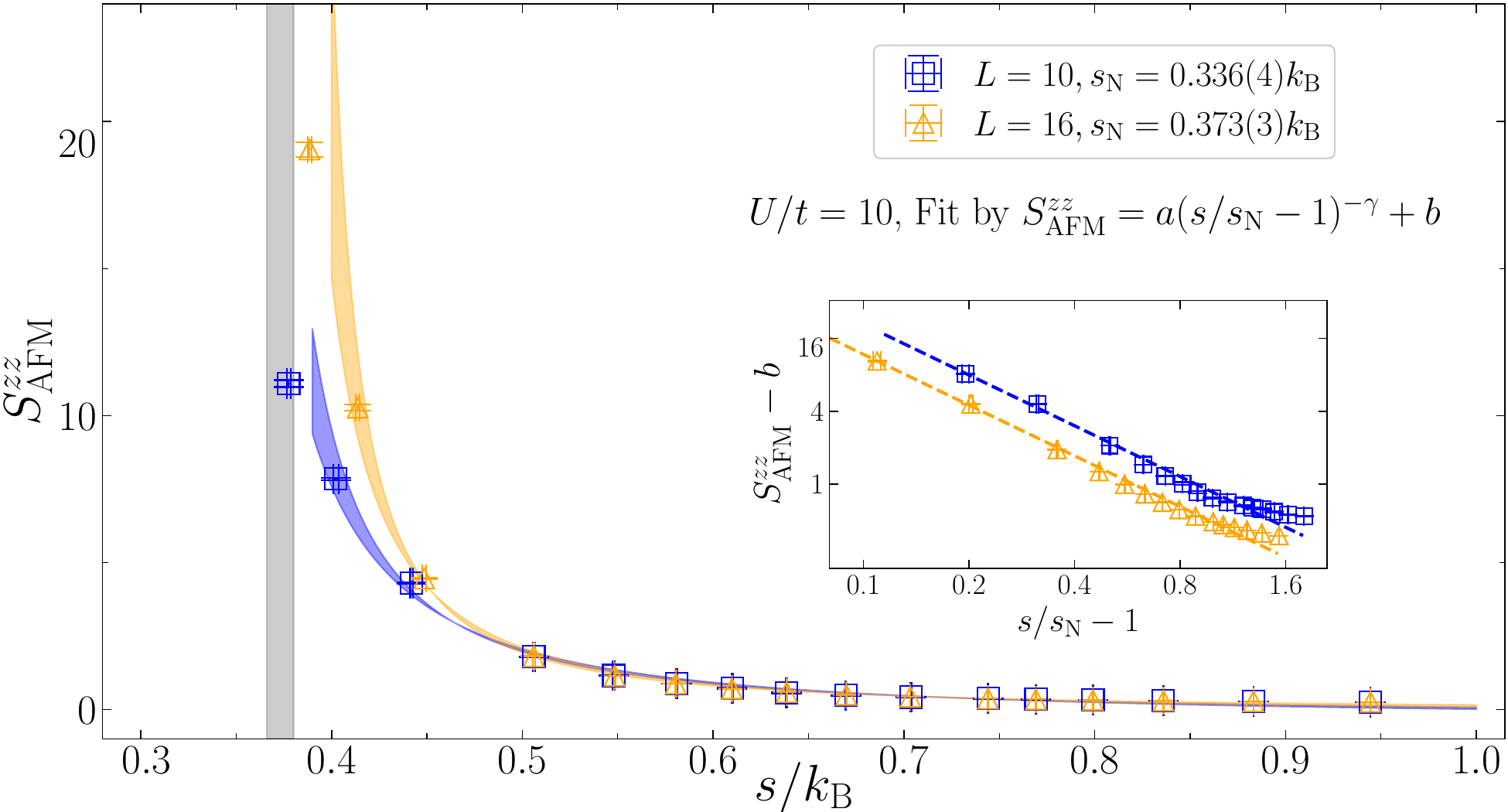}
\caption{\label{fig:FigS2Scaling} Critical scaling of $S_{\rm AFM}^{zz}$ versus the entropy per particle $\boldsymbol{s}/k_B$ for $U/t=10$ with $L=10$ and $L=16$, using the relation $S_{\rm AFM}^{zz}=a(\boldsymbol{s}/\boldsymbol{s}_{\rm N}-1)^{-\gamma}+b$ with $\gamma=1.396$ as the critical exponent from the Heisenberg universality class~\cite{Campostrini2002}. The inset is the semi-log plot of $(S_{\rm AFM}^{zz}-b)$ versus $(\boldsymbol{s}/\boldsymbol{s}_{\rm N}-1)$, with dashed lines denoting the fitting curves.}
\end{figure}

\section{The critical scaling of AFM structure factor versus entropy}
\label{sec:Scaling}

In Fig.~\ref{fig:FigS2Scaling}, we present the critical scaling of AFM structure factor $S_{\rm AFM}^{zz}$ versus the thermal entropy per particle $\boldsymbol{s}$ for $U/t=10$ with $L=10$ and $L=16$ systems. We adopt the formula $S_{\rm AFM}^{zz}=a(\boldsymbol{s}/\boldsymbol{s}_{\rm N}-1)^{-\gamma}+b$ for the fitting using the numerical data in PM phase, following the manner in the experiment~\cite{Shao2024}. The results indeed fits the scaling relation quite well, and the critical extropy $\boldsymbol{s}_{\rm N}/k_B$ from the fitting is comparable to numbers presented in the main text. We also note that the critical extropy extracted from the experimental data~\cite{Shao2024} is significantly smaller than our numerical results, such as $\boldsymbol{s}_{\rm N}/k_B=0.27(1)$ for $U/t=11.75$ in experiment comparing to $\boldsymbol{s}_{\rm N}/k_B=0.33(1)$ for $U/t=10$ in our numerics. This disagreement might be attributed to two reasons. First, the experiment data process used the initial single-particle entropy (before loading the optical lattice and adding the interaction), and the experimental measurements with increasing $U$ should have entropy increase as discussed in the main text. Second, the existence of lattice density disorder in the experimental setup should also suppress the N\'{e}el transition temperatures of the system, which consequently reduces the critical entropy.

\section{The raw data of the thermal entropy}

In this section we present the thermal entropy data underlying Fig.~1 of the main text. The entropy is obtained at fixed temperatures using the method introduced in Ref.~\cite{Song2024b}, by numerically integrating the double occupancy as a function of interaction. The data cover a large parameter space, the temperature ranging from $T=0.2t$ to $0.7t$ and interaction strength from $U=t$ to $13t$. For $T \leq 0.4t$, the results are obtained from the systems with $L=12$, while for $T>0.4t$, a smaller lattice with $L=8$ is used, where finite-size effects are less pronounced and are verified to be negligible. These results may serve as benchmarks for future theoretical studies as well as optical lattice experiments.

\begin{table}[h]
\caption{The thermal entropy per site of the three-dimensional half-filled Fermi Hubbard model on the simple-cubic lattice over a broad parameter range. The entropy is evaluated at fixed temperatures by numerically integrating the double occupancy as a function of interaction ~\cite{Song2024b}. For $T/t \leq 0.40$, the results are obtained from the systems with $L=12$, while for $T/t > 0.40$ a smaller system size $L=8$ is used, where finite-size effects are less pronounced and are verified to be negligible.}
\begin{tabular}{|c|c|c|c|c|c|c|c|c|c|c|c|}
\hline
\diagbox{$U/t$}{$T/t$} & 0.200 &0.250&0.290&0.335&0.360&0.400&0.450&0.500&0.550&0.600&0.700\\
\hline
1 &0.209(7) &0.2454(2) &0.2792(1) &0.31962(7) &0.3417(2) &0.3774(1) &0.4266(1) &0.4675(1) &0.5073(2) &0.5460(2) &0.6188(1)\\
\hline
2 &0.213(6) &0.2538(4) &0.2897(2) &0.33095(6) &0.35355(8) &0.38974(7) &0.43817(9) &0.4781(1) &0.5180(2) &0.5557(2) &0.62654(6)\\
\hline
3 &0.226(6) &0.2666(6) &0.3039(3) &0.3480(1) &0.3717(2) &0.4083(1) &0.4557(1) &0.4952(1) &0.5343(2) &0.5707(1) &0.6381(1)\\
4 &0.226(6) &0.2861(6) &0.3281(4) &0.3711(1) &0.3957(2) &0.4320(1) &0.4777(1) &0.5168(2) &0.5539(2) &0.5886(3) &0.6512(1)\\
\hline
5 &0.098(6) &0.2623(8) &0.3466(5) &0.3992(2) &0.4239(2) &0.4597(2) &0.5030(3) &0.5407(2) &0.5748(3) &0.6069(2) &0.6636(1)\\
\hline
6 &0.049(6) &0.144(2)  &0.282(1)  &0.4110(4) &0.4434(4) &0.4829(2) &0.5256(2) &0.5620(2) &0.5938(4) &0.6232(2) &0.6736(2)\\
\hline
7 &0.040(6) &0.105(4)  &0.210(1)  &0.3808(6) &0.4429(4) &0.4917(3) &0.5381(5) &0.5745(3) &0.6062(4) &0.6336(3) &0.6782(2)\\
\hline
8 &0.043(6) &0.108(2)  &0.199(1)  &0.3540(8) &0.4334(9) &0.4897(3) &0.5402(2) &0.5775(2) &0.6086(4) &0.6347(2) &0.6753(2)\\
\hline
9 &0.047(7) &0.127(2)  &0.224(2)  &0.3719(7) &0.4364(6) &0.4893(4) &0.5402(5) &0.5766(3) &0.6056(2) &0.6304(2) &0.6679(2)\\
\hline
10&0.056(7) &0.153(2)  &0.266(2)  &0.4070(6) &0.4551(7) &0.4978(4) &0.5446(4) &0.5771(3) &0.6031(4) &0.6260(3) &0.6598(2)\\
\hline
11&0.073(6) &0.186(1)  &0.324(1)  &0.4370(6) &0.4794(5) &0.5116(4) &0.5540(4) &0.5822(3) &0.6048(2) &0.6248(2) &0.6537(2)\\
\hline
12&0.095(6) &0.230(2)  &0.384(2)  &0.462(1)  &0.5017(7) &0.5273(4) &0.5654(4) &0.5898(4) &0.6094(5) &0.6263(3) &0.6502(2)\\
\hline
13&0.119(6) &0.293(3)  &0.422(3)  &0.489(1)  &0.5197(8) &0.5428(6) &0.5774(6) &0.5991(5) &0.6164(3) &0.6303(4) &0.6499(3)\\
\hline
\end{tabular}
\end{table}

\end{document}